\documentclass[pra,onecolumn,superscriptaddress,showpacs]{revtex4}
\usepackage{amsmath}
\usepackage{graphicx}
\usepackage{bm}
\usepackage{amsbsy}
\usepackage{amsfonts}
\usepackage{amsthm}

\newcommand{\nn}{\nonumber \\}

\newcommand{\dt}{\Delta \lambda}
\newcommand{\initt}{\mu}
\newcommand{\nbound}{2m5^{k-1}\left\lceil 5k\Lambda\dt \left(\frac{5}{3} \right)^{k}\left(\frac{\Lambda\dt}{\epsilon}\right)^{1/2k}\right\rceil}

\newcommand{\ubound}{\max_{x>y}\|U(x,y)\|}

\def\squareforqed{\hbox{\rlap{$\sqcap$}$\sqcup$}}
\def\qed{\ifmmode\squareforqed\else{\unskip\nobreak\hfil
\penalty50\hskip1em\null\nobreak\hfil\squareforqed
\parfillskip=0pt\finalhyphendemerits=0\endgraf}\fi}

\newtheorem{theorem}{Theorem}
\newtheorem{lemma}{Lemma}
\newtheorem{definition}{Definition}
\newtheorem{corollary}{Corollary}

\newenvironment{proofb}{\begin{trivlist}\item[]{\flushleft\it Proof. }}
{\qed\end{trivlist}}
\newenvironment{proofof}[1]{\begin{trivlist}\item[]{\flushleft\it
Proof of~#1. }}
{\qed\end{trivlist}}

\begin{document}
\title{Higher Order Decompositions of Ordered Operator Exponentials}
\begin{abstract}
We present a decomposition scheme based on Lie-Trotter-Suzuki product
formulae to represent an ordered operator exponential as a product of
ordinary operator exponentials. We provide a rigorous proof that does not
use a time-displacement superoperator, and can be applied to
non-analytic functions. Our proof provides explicit bounds on the error
and includes cases where the functions are not infinitely differentiable.
We show that Lie-Trotter-Suzuki product formulae can still be used for
functions that are not infinitely differentiable, but that arbitrary order
scaling may not be achieved.
\end{abstract}
\author{{{Nathan Wiebe}}}
\affiliation{\textit{Institute for Quantum Information Science, University
of Calgary,
Alberta T2N 1N4, Canada.}}
\author{Dominic Berry}
\affiliation{\textit{Centre for Quantum Computer Technology, Macquarie
University, Sydney, NSW 2109, Australia.}}
\author{\text{Peter H\o yer}}
\affiliation{\textit{Institute for Quantum Information Science, University
of Calgary,
Alberta T2N 1N4, Canada.}}
\affiliation{\textit{Department of Computer Science, University of
Calgary, Alberta T2N 1N4, Canada.}}

\author{Barry C. Sanders}
\affiliation{\textit{Institute for Quantum Information Science, University
of Calgary,
Alberta T2N 1N4, Canada.}}
\maketitle

\section{Introduction}\label{sec:intro}
Decompositions of operator exponentials are widely used to approximate
operator
exponentials that arise in both physics and applied mathematics.
These approximations are used because it is often difficult to
exponentiate an operator directly,
even if the operator is a sum of a sequence of operators that can be
easily exponentiated individually.
The goal in a decomposition method is to approximate an exponential of a
sum of operators as a product of operator
exponentials.
Particular examples of decompositions include the Trotter formula, and the
related Baker-Campbell-Hausdorff formula as well as Suzuki decompositions
\cite{suzuki:physa,suzuki:mathphys}.  These approximations have found use
in many fields including quantum Monte-Carlo calculations
\cite{suzuki:physa}, quantum computing \cite{nielsenchuang} and classical
dynamics \cite{chin:integrator} to name a few.

The problem is to solve for the operator $U$ given by
\begin{equation}\label{eq:udiffeq}
\partial_{\lambda}U(\lambda,\mu)=H(\lambda)U(\lambda,\mu),
\qquad\textnormal{with } U(\mu,\mu)=\openone,
\end{equation}
where $H$ is a linear operator and $\lambda$ and $\mu$ are real numbers.
In general, computing~$U$ can be difficult regardless of whether~$H$ is
dependent on $\lambda$. The case of $\lambda$-dependence makes the problem
significantly more complicated and is the focus of this work, whereas the
case of $\lambda$-independent $H$ has been well studied. The Lie-Trotter
formula gives a simple solution, whereas more efficient higher-order
solutions are given by the Lie-Trotter-Suzuki (LTS) product
formulae~\cite{suzuki:mathphys,sanderssim}.

A direct approach to solving~$U$ in the $\lambda$-independent case is
first to diagonalize~$H$, then solve the differential
equation~(\ref{eq:udiffeq}) by direct exponentiation. The scenario we
consider is that this is not possible, and instead we are given a set of
$m$ operators $\{H_j:j=1,\ldots,m\}$, where it is possible to
exponentiate of each of these operators, and
\begin{equation}
\label{eq:Hsum}
    H=\sum_{j=1}^m H_j.
\end{equation}
The evolution operator can then be approximated by a product of
exponentials of $H_j$ for some sequence of $\{j_i\}$ and intervals
$\{\Delta\lambda_i\}$,
\begin{equation}
\label{eq:Uprod}
    U(\lambda,\mu)\approx\exp(H_{j_N}\Delta \lambda_N)\exp(H_{j_{N-1}}\Delta \lambda_{N-1})\cdots\exp(H_{j_1}\Delta \lambda_1)=\prod_{i=1}^N e^{H_{j_i}\Delta \lambda_i}.
\end{equation}
The goal is to make an intractable calculation of~$U$ tractable by
approximating $U$ as a finite-length product of efficiently calculated
exponentials. The complexity of the calculation can then be quantified by
the number of exponentials $N$, and the scaling of $N$ in terms of the
parameter difference $\Delta\lambda=\lambda-\mu$.

The case of $\lambda$-dependence makes the problem harder because, rather
than the usual operator exponential of~$H$, the solution is an ordered
exponential. Diagonalization techniques are
not directly applicable to solving ordered exponentials, and methods such
as using the ordered product of exponentials are needed. That is, $U$ is
approximated by an expression of the form
\begin{equation}
\label{eq:Uprod2}
    U(\lambda,\mu)\approx\prod_{i=1}^N e^{H(\lambda_i)\Delta \lambda_i}.
\end{equation}

{We consider the case where $H$ may be $\lambda$-dependent
\emph{and} where $H$ is a sum as in \eqref{eq:Hsum}}. One approach would be to replace
Eq.~\eqref{eq:Uprod} with a product formula of ordered exponentials. There
would still remain the problem of evaluating the ordered exponentials,
which would require an approach such as \eqref{eq:Uprod2}. A simpler
approach is to use \eqref{eq:Uprod}, but choose appropriate values of
$\lambda$ at which to evaluate the $H_{j_i}$. The approximation is then
\begin{equation}
\label{eq:UjiHji}
    U(\lambda,\mu)\approx\prod_{i=1}^N e^{H_{j_i}(\lambda_i)\Delta
\lambda_i}.
\end{equation}

Suzuki provides an efficient method for approximating ordered exponentials
in this way \cite{suzuki:mathphys}. The method given by Suzuki is in terms
of a time-displacement operator, but is equivalent. For many applications
it is desirable to be able to place upper bounds on the error that can be
obtained. Suzuki derives an order scaling, but not an upper bound on the
error. Berry {et al.}\ find an upper bound on the error, but only
in the case without $\lambda$-dependence, and for $H$ antihermitian
(corresponding to Hamiltonian evolution) \cite{sanderssim}.

Here we prove upper bounds on the error in the general case where $H$ can
depend on $\lambda$, and is not restricted to be antihermitian. Whereas
Suzuki uses a time-displacement operator, we provide a proof entirely
without the use of this operator. The time-displacement operator is
problematic, because it is unclear how it acts for $\lambda$-dependence
that is non-analytic. We find that, provided all derivatives of the
$H_j(\lambda)$ exist, Suzuki's result holds. If there are derivatives that
do not exist, then Suzuki's result does not necessarily hold; we
demonstrate this via a counterexample. We solve the case where derivatives
may not exist, and find that Suzuki's approach can still give better
scaling of $N$ with $\Delta\lambda$, although the scaling that can be
obtained is limited by how many times the $H_j(\lambda)$ are
differentiable.

In Sec.\ \ref{sec:decomp} we give the background for Trotter product
formulae in detail. In Sec.\ \ref{sec:Suzuki} we review Suzuki's
decomposition methods, and provide our form of Suzuki's recursive method.
In Sec.\ \ref{sec:sufficient} we introduce our terminology and present our
main result. Then we rigorously prove the scaling of the error in Sec.\
\ref{sec:suzukianalytic}, and place an upper bound on the error in Sec.\
\ref{sec:suzerror}.  We then use the error bounds in Sec.\ \ref{sec:suplin} to
find the appropriate order of the integrator to use.

\section{Trotter formulae}\label{sec:decomp}
Typically there are two different scenarios that may be considered. First,
one may consider a short interval $\Delta \lambda$; the goal is then to
obtain error that decreases rapidly as $\Delta \lambda\to 0$. Alternatively
the interval $\Delta \lambda$ may be long, and the goal is to obtain an
approximation to within a certain error with as few exponentials as
possible. For example, given $\lambda$-independent operators $A$ and $B$,
it holds that
\begin{equation}
\label{eq:short}
e^{\Delta \lambda (A+B)} = e^{\Delta \lambda A}e^{\Delta \lambda
B}+O(\Delta \lambda^2).
\end{equation}
This gives an accurate approximation for small $\Delta \lambda$. For large
$\Delta \lambda$, we may use Eq.~\eqref{eq:short} to derive the Trotter
formula
\begin{equation}
\label{eq:long}
e^{\Delta \lambda (A+B)} = (e^{\Delta \lambda A/n}e^{\Delta \lambda
B/n})^n+O(\Delta \lambda^2/n).
\end{equation}
This order of error is obtained because the error for interval $\Delta
\lambda/n$ is $O((\Delta \lambda/n)^2)$. Taking the power of $n$ then
gives $n$ times this error {if the
norm of $\exp[(A+B)\dt]$ is at most one for any $\dt>0$}, resulting in the error shown in
Eq.~\eqref{eq:long}. To obtain a given error $\epsilon$, the value of $n$
must then scale as $O(\Delta \lambda^2/\epsilon)$. The goal is to make the
value of $n$ needed to achieve a given accuracy as small as possible ($n$
is proportional to the total number of exponentials).

More generally, for a sum of an arbitrary number of operators $H_j$,
similar formulae give the same scaling. To obtain better scaling, one can
use a different product of exponentials. The Lie-Trotter-Suzuki product
formulae \cite{suzuki:mathphys} replace the product for short
$\Delta\lambda$ with another that gives error scaling as $O(\Delta
\lambda^{p+1})$.

It can be seen that splitting
large $\Delta \lambda$ into $n$ intervals as in Eq.~\eqref{eq:long} yields
an error scaling as $O(\Delta \lambda^{p+1}/n^{p})$ if the norm of $U$ is
at most one for any $\lambda$. It may at first
appear that this gives \emph{worse} results for large $\Delta\lambda$ due
to the higher power. In fact, there is an advantage due to the fact that a
higher power of $n$ is obtained. The value of $n$ required to achieve a
given error then scales as $O(\Delta\lambda^{1+1/p}/ \epsilon^{1/p})$.
Therefore, for large $\Delta\lambda$, increasing $p$ gives scaling of $N$
that is close to linear in $\Delta\lambda$.

Similar considerations hold for the case of ordered exponentials (i.e.\
with $\lambda$-dependence).  Huyghebaert and De Raedt showed how
to generalize the Trotter formula to apply to ordered operator
exponentials \cite{huyghebaert:timedeptrotter}.  Their formula has a decomposition error that is
$O(\dt^{2})$, but requires that the integrals of $A(u)$ and $B(u)$ are known.
Subsequently Suzuki developed a method to achieve error
that scales as $O(\Delta \lambda^{p+1})$ for some ordered exponentials
\cite{suzuki:jacad}, and does not require the integrals of $A$ and
$B$ to be known. We find that, in contrast to the
$\lambda$-independent case, it is not necessarily possible to obtain
scaling as $O(\Delta \lambda^{p+1})$ for arbitrarily large $p$. It is
possible if derivatives of all orders exist. If there are higher-order
derivatives that do not exist, then it is still possible to use Suzuki's
method to obtain error scaling as $O(\Delta \lambda^{p+1})$ for some
values of $p$, but the maximum value of $p$ for which this scaling can be
proven depends on what orders of derivatives exist.

\section{Suzuki Decompositions}\label{sec:Suzuki}
In this section we explain Suzuki decompositions in more detail. In general, decompositions are of the form, as in \eqref{eq:UjiHji},
\begin{equation}
\label{eq:Up}
    \tilde U(\initt+\dt,\initt)=\prod_{i=1}^N e^{H_{j_i}(\lambda_i)\Delta\lambda_i}.
\end{equation}
Here we write the final parameter $\lambda$ as $\initt+\dt$, to emphasize the dependence on $\dt$ ($=\lambda-\initt$).
There are many different types of decompositions, but the type that we focus on in this paper is symmetric decompositions because all Suzuki decompositions are symmetric.
\begin{definition}\label{def:symmetric}
The operator $\tilde U(\initt+\dt,\initt)$ is a symmetric decomposition of the operator $U(\initt+\dt,\initt)$ if $\tilde U(\initt+\dt,\initt)$ is a decomposition of $U(\initt+\dt,\initt)$ and $\tilde U(\initt+\dt,\initt)=[\tilde U(\initt,\initt+\dt)]^{-1}$.
\end{definition}

An important method for generating symmetric decompositions
is due to Suzuki \cite{suzuki:timeorderbook,suzuki:jacad}, which we call Suzuki's recursive method due to its similarity to the method presented by Suzuki in \cite{suzuki:mathphys}.  Furthermore we call any decomposition formula that is found using this method a Suzuki decomposition.

\label{def:suzdecomposition}
 Suzuki's recursive method takes a symmetric decomposition formula $U_p(\initt+\dt,\initt)$, that approximates an ordered operator exponential $U(\initt+\dt,\initt)$  with an approximation error  that is at most proportional to $\dt^{2p+1}$ as input, and outputs a symmetric approximation formula $U_{p+1}(\initt+\dt,\initt)$ with an error that is often proportional to $\dt^{2p+3}$.  The approximation $U_{p+1}(\initt+\dt,\initt)$  is
found using the following recursion relations,
\begin{align}
U_{p+1}(\initt+\dt,\initt)&\equiv U_p(\initt+\dt,\initt+[1-s_p]\dt)U_p(\initt+[1-s_p]\dt,\initt+[1-2s_p]\dt)\nonumber\\&\times U_p(\initt+[1-2s_p]\dt,\initt+2s_p\dt)
U_p(\initt+2s_p\dt,\initt+s_p\dt)U_p(\initt+s_p\dt,\initt),
\end{align}
with $s_p\equiv \left(4-4^{1/(2p+1)}\right)^{-1}$.

Suzuki's recursive method does not actually approximate $U(\initt+\dt,\initt)$ but rather it builds
a higher order approximation formula out of a lower order one.  Therefore this method can only be used to approximate $U(\initt+\dt,\initt)$ if it is seeded with an appropriate initial approximation.  A convenient approximation formula based on Suzuki's recursive method is the $k^{\text{th}}$ order Lie-Trotter-Suzuki product formula which is defined as follows.
\begin{definition}\label{def:lietrotsuz}
The $k^{\text{th}}$ order Lie-Trotter-Suzuki product formula for the operator $H(u)=\sum_{j=1}^m H_j(u)$ and the interval $[\initt,\initt+\dt]$ is defined to be $U_k(\initt+\dt,\initt)$, which is
found by using
\begin{align}\label{eq:lietrot}
U_1(\initt+\dt,\initt)&\equiv\left(\prod_{j=1}^m\exp({H}_j(\initt+\dt/2)\dt/2)
\right)\left(\prod_{j=m}^1\exp({H}_j(\initt+\dt/2)\dt/2)\right),
\end{align}
as an initial approximation and by applying Suzuki's recursive method to it $k-1$ times.
\end{definition}

Based on Suzuki's analysis \cite{suzuki:timeorderbook,suzuki:jacad} $U_k$ should have approximation error that is proportional to $\dt^{2k+1}$. Hence if $\dt$ is sufficiently small, then the formula should be highly accurate. One might think that it would be advantageous to increase $k$ without limit, in order to obtain increasingly accurate approximation formulae. This is not the case, because the number of terms in the formula increases exponentially with $k$. The best value of $k$ to use can be expected to depend on the desired accuracy, as well as a range of other parameters \cite{sanderssim}.

\section{Sufficiency Criterion for Decomposition}\label{sec:sufficient}
Suzuki's recursive method is a powerful technique for generating high-order decomposition formulae for ordered operator exponentials.  The $k^{\text{th}}$ order Lie-Trotter-Suzuki product formula in particular seems to be well suited for approximating ordered operator exponentials that appear in quantum mechanics and in other fields; furthermore it appears that these formulae should be applicable to approximating the ordered exponentials of any finite dimensional operator $H$.  However it turns out that Suzuki's recursive method does not always generate a higher order decomposition formula from a lower order one.

\begin{figure}[t!]
\centering

{\includegraphics[width=1.0\textwidth]{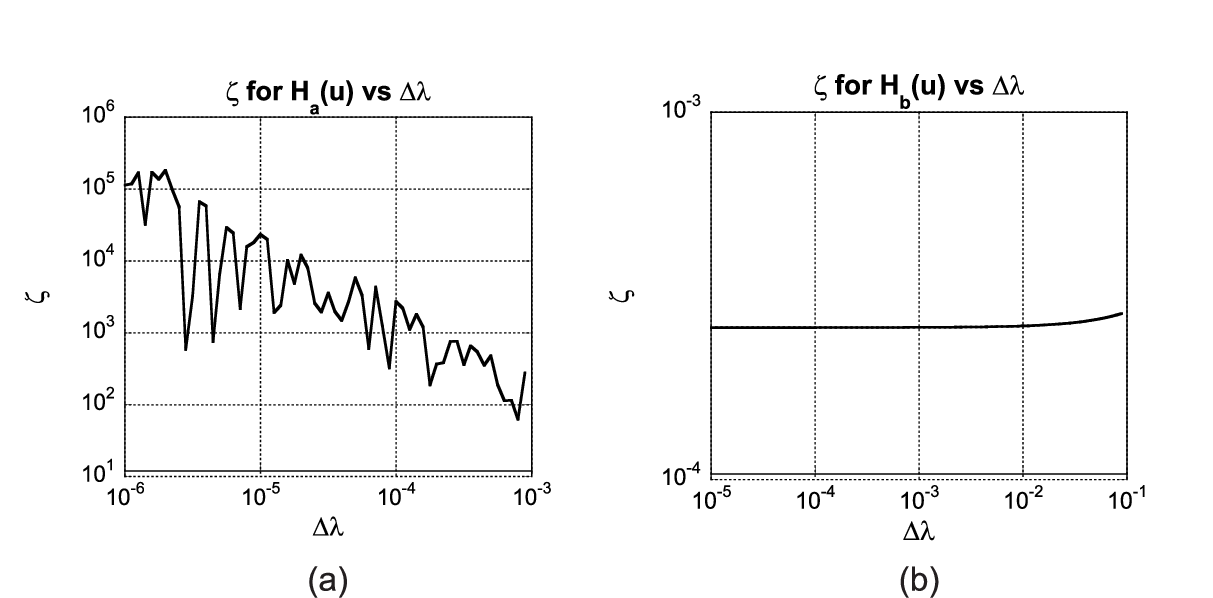}}
\caption{This is a plot of $\zeta=\|U(\dt,0)-U_2(\dt,0)\|_2/\dt^5$ for $H_a=u^3\sin(1/u)\openone$ in (a) and $H_b=\cos(u)\openone$ in (b).  The error in (a) is proportional to $\dt^4$ as opposed to the $O(\dt)^5$ scaling predicted for that Suzuki decomposition.  The error in (b) is proportional to $\dt^5$ as expected for that Suzuki decomposition.\label{fig:errorplot}}
\end{figure}

We show this using the example of the operator $H_2(u)=u^3\sin(1/u)\openone$. For this operator the second order Lie-Trotter-Suzuki product formula is not an approximation whose error as measured by the 2-norm is $O(\dt^{5})$. In Figure \ref{fig:errorplot} we see that the error is proportional to $\dt^{4}$ for the operator
$H_a(u)$, rather than the $\dt^5$ scaling that we expect and observe for the analytic operator $H_b(u)=\cos(u)$.  This shows that the second order Lie-Trotter-Suzuki product formula is not as accurate as may be expected for some non-analytic operators.

Our analysis will show that this discrepancy arises from the fact that $H_2(u)=u^3\sin(1/u)$ is not smooth enough for the second order Lie-Trotter-Suzuki formula to have an error which is $O(\dt^{5})$.  In the subsequent discussion we will need to classify the smoothness of the operators that arise in decompositions.  We use the smoothness criteria $2k$-smooth and $\Lambda$-$2k$-smooth, which we define below.
\begin{definition}\label{def:suzsmooth}
The set of operators $\{H_j:j=1,\dots, m\}$ is $P$-smooth on the interval~$[\initt,\lambda]$ if for each $H_j$ the quantity $\|\partial_u^{P}H_j(u)\|$ is finite on the interval~$[\initt,\lambda]$.
\end{definition}
{Here, and throughout this paper, we define $\|\cdot\|$ to be the 2-norm.  Also if $\{H_j\}$ is $P$-smooth for every positive integer $P$, we call $\{H_j\}$ $\infty$-smooth.}

{This condition is not precise enough for all of our purposes. For our error bounds we need to introduce the more
precise condition of $\Lambda$-$P$-smoothness.  This condition is useful because it guarantees that if the set $\{H_j\}$ is $\Lambda$-$P$-smooth and $p\leq P$ then $\|H^{(p)}(u)\|\leq \Lambda^p$.  This property allows us to write our error bounds in a form that does not contain any of the derivatives of $H$ individually, but rather in terms of $\Lambda$ which upper bounds the magnitude of any of these derivatives.  We formally define this condition below.}
\begin{definition}\label{def:lambdasuzsmooth}
The set of operators $\{H_j:j=1,\dots, m\}$ is $\Lambda$-$P$-smooth on the interval~$[\initt,\lambda]$ if $\{H_j\}$ is $P$-smooth and~$\Lambda\geq\sup_{p=0,1,...,P}\left(\sup_{u\in[\initt,\lambda]}\left(\sum_{j=1}^m\|H_j^{(p)}(u)\|\right)^{1/(p+1)} \right)$.
\end{definition}

For example if $\{H(u)\}=\{\sin(2u)\openone\}$, where $\openone$ is the identity operator, then using Definition \ref{def:lambdasuzsmooth} $\{H(u)\}$ is $2^{2/3}$-$2$-smooth on the interval $[0,\pi]$ because the largest value $\|H(u)^{(p)}\|^{1/(p+1)}$ takes is $2^{2/3}$, for $p=0,1,2$.  It is also $2$-$2$-smooth because $2^{2/3}\leq 2$, furthermore since $\|H(u)^{(p)}\|^{1/(p+1)}< {2}$ for all positive integers $p$ then $\{H(u)\}$ is also ${2}$-$\infty$-smooth.

Using this measure of smoothness we can then state the following theorem, which is also the main theorem in this paper.
\begin{theorem}\label{thm:nexp}
If the set $\{H_j\}$ is $\Lambda$-$2k$-smooth on the interval $[\initt,\initt+\dt]$, and $\epsilon\leq (9/10)(5/3)^{k}\Lambda\dt$, and $\epsilon\leq 1$ and \mbox{$\ubound\le 1$}, then a decomposition $\tilde{U}(\initt+\dt,\initt)$ can be constructed such that $\|\tilde{U}-U\|\le\epsilon$ and the number of operator exponentials present in $\tilde{U}$, $N$, satisfies
\begin{equation}
N\leq \nbound.
\end{equation}
\end{theorem}

We prove Theorem \ref{thm:nexp} in several steps, the details of which are spread over Secs.~\ref{sec:suzukianalytic} and \ref{sec:suzerror}.  In Sec.~\ref{sec:suzukianalytic} we construct the Taylor series for an ordered operator exponential, and use this series to prove that the Lie-Trotter-Suzuki product formula can generate an approximation whose error is $O(\dt^{2k+1})$, if $\{H_j\}$ is $2k$-smooth on the interval $[\initt,\initt+\dt]$.  In Sec.~\ref{sec:suzerror} we use the order estimates in Sec.~\ref{sec:suzukianalytic} to obtain upper bounds on the error.  The result of Theorem \ref{thm:nexp} then follows by counting the number of exponentials needed to make the error bound less than $\epsilon$.  Finally in Sec.~\ref{sec:suplin} we show that if $k$ is chosen appropriately, then $N$ scales almost optimally with $\dt$ if {there exists a value of $\Lambda$ such that $\{H_j\}$ is $\Lambda$-$\infty$-smooth on $[\initt,\initt +\dt]$ for every $\dt>0$}.
\section{Decomposing Ordered Exponentials}\label{sec:suzukianalytic}

In this section we present a new derivation of Suzuki's recursive method.  Our derivation has the advantage that it can be rigorously proven that if $\{H_j\}$ is $2k$-smooth, where $H(u)=\sum_{j=1}^m H_j(u)$, then the $k^{\text{th}}$ order Lie-Trotter-Suzuki product formula will have an error of $O(\dt^{2k+1})$.  We show this in three steps.  We first give an expression for the Taylor series expansion of a ordered exponential $U(\initt+\dt,\initt)$.  Then using this expression for the Taylor series, we show in Theorem \ref{thm:suzgeneral} that Suzuki's recursive method can be used to generate approximations to $U(\initt+\dt,\initt)$ that invoke an error that is $O(\dt^{2k+1})$ if $\{H_j\}$ is $2k$-smooth.  Finally we show in Corollary \ref{cor:lietrotsuz} that if $\{H_j\}$ is $2k$-smooth then the  $k^{\text{th}}$ order Lie-Trotter-Suzuki product formula has an error that is at most proportional to $\dt^{2k+1}$.

It is convenient to expand $U$ in a Taylor series of the form
\begin{equation}
U(\initt+\dt,\initt)=\openone + T_1(\initt)\dt+\frac{T_2(\initt)\dt^2}{2!}+\cdots.
\end{equation}
If $H$ is not analytic, then this Taylor series must be truncated, and the error can be bounded by the following lemma.
\begin{lemma}\label{lem:taylr}
For all $H:\mathbb{R}\rightarrow \mathbb{C}^{N\times N}$, {that are for}
$P\in\mathbb{N}_0$, $P$ times differentiable on the interval $[\initt,\initt+\dt]\subset \mathbb{R}$, then
\begin{align}
\left\|U(\initt+\dt,\initt)-\sum_{p=0}^P\frac{(\dt)^pT_p}{p!}\right\|&\leq
\frac{\max_{u\in [\initt,\initt+\dt]}\|T_{P+1}(u)U(u,\initt)\|\dt^{P+1}}{(P+1)!},
\end{align}
where $T_p(u)$ is defined by the recursion relation, $T_{p+1}(u)\equiv
T_p(u) H(u)+\partial_tT_p(u)$, with $T_0\equiv\openone$ chosen to be the initial condition.
\end{lemma}

\begin{proofb}

We first show, for the positive integer $\ell\le P+1$ that if $s\in[\initt,\initt+\dt]$ then
\begin{equation}\label{eq:tp3}
\frac{\partial^\ell}{\partial s^\ell}U(s,\initt) = T_\ell(s)U(s,\initt).
\end{equation}
This equation can be validated by using induction on $\ell$.
The base case follows by setting $\ell=0$ in \eqref{eq:tp3}. We then demonstrate the induction step by noting that if \eqref{eq:tp3} is true for $\ell\le P$ then
\begin{align}
\frac{\partial^{\ell+1}}{\partial s^{\ell+1}}U(s,\initt)&=\frac{\partial}{\partial s} T_\ell(s)U(s,\initt) \nn
&=\left[\frac{\partial}{\partial s}T_\ell(s)\right]U(s,\initt)+ T_\ell(s)\left[\frac{\partial}{\partial s}U(s,\initt)\right]\label{eq:diffeq}.
\end{align}
Since $T_\ell(s)$ contains derivatives of $H(s)$ up to order $\ell-1$, it follows that
$\partial_sT_\ell(s)$ contains derivatives up to order $\ell$. Then since $H(s)$
is $P$ times differentiable, $\partial_sT_\ell(s)$ exists if $\ell\leq P$,  which implies that
 $\partial_s^{\ell+1}U(s,\initt)$ exists.  We then use the differential equation in \eqref{eq:udiffeq} to evaluate the derivative of $U(\initt+\dt,\initt)$ in \eqref{eq:diffeq} and use the fact that $T_{\ell+1}=HT_{\ell}+\partial_s T_{\ell}$ to find that
\begin{align}
\frac{\partial^{\ell+1}}{\partial s^{\ell+1}}U(s,\initt)&=T_{\ell+1}(s)U(s,\initt).
\end{align}
This demonstrates the induction step in our proof of \eqref{eq:tp3}.  Since we have already shown that \eqref{eq:tp3} is valid for $T_0$, it is also true for all $T_\ell$ if $\ell\le P+1$ by induction on $\ell$.

We then use \eqref{eq:tp3} and Taylor's Theorem to conclude that
\begin{equation}\label{eq:dominic2}
U(\initt+\dt,\initt)=\sum_{p=0}^P\frac{(\dt)^pT_p(\initt)}{p!}+\int_0^{\dt}T_{P+1}(\initt+s)U(\initt+s,\initt)\frac{(\dt-s)^P}{P!}\mathrm{d}s.
\end{equation}
We rearrange this result and find that
\begin{align}
\left\|U(\initt+\dt,\initt)-\sum_{p=0}^P\frac{(\dt)^pT_p}{p!}\right\|
&\le \frac{\max_{u\in [\initt,\initt+\dt]}\|T_{P+1}(u)U(u,\initt)\|\dt^{P+1}}{(P+1)!}.
\end{align}
\end{proofb}

Lemma \ref{lem:taylr} provides a convenient expression for the terms in the Taylor series of $U(\initt+\dt,\initt)$, and it also estimates the error invoked by truncating the series at order $P$ for any $P\in \mathbb{N}$.  The
following theorem uses this Lemma to show that  Suzuki's recursive method will produce a higher order
approximation from a lower order symmetric approximation, if $\{H_j\}$ is sufficiently smooth on $[\initt,\initt+\dt]$ and if $H$ is the sum of all of the elements in the set $\{H_j\}$.

\begin{theorem}\label{thm:suzgeneral}
If $H=\sum_{j=1}^m H_j$ where the set $\{H_j\}$ is for a fixed $p$, $2(p+1)$-smooth on the interval $[\initt,\initt+\dt]$ and $U_p(\initt+\dt,\initt)$ is a symmetric approximation formula such that
$\|U_p(\initt+\dt,\initt)-U(\initt+\dt,\initt)\|\in O(\dt^{2p+1})$, and $U_{p+1}(\initt+\dt,\initt)$ is found
by applying Suzuki's recursive method on $U_p(\initt+\dt,\initt)$, then
\begin{equation}
\|U(\initt+\dt,\initt)-U_{p+1}(\initt+\dt,\initt)\|\in O(\dt^{2p+3}).
\end{equation}
\end{theorem}

\begin{proofb}

In this proof we compare the Taylor series of $U$ to that of $U_p$ and show that by choosing $s_p$ appropriately will cause both the
terms proportional to $\dt^{2p+2}$ and $\dt^{2p+3}$ to vanish.

By expanding the recursive formula in Lemma \ref{lem:taylr}
we see that a Taylor polynomial can be constructed for $U$ whose difference from $U$ is $O(\dt^{2p+3})$
because $\{H_j\}$ is $2(p+1)$-smooth on $[\initt,\initt+\dt]$.  A similar polynomial can be constructed for $U_p$ by
Taylor expanding each $H_j$ that appears in the exponentials in $U_p$, and then expanding each of these exponentials.
Then because $\{H_j\}$ is $2(p+1)$-smooth, Taylor's Theorem implies that this polynomial can be constructed such that the
difference between it
 and $U_p$ is $O(\dt^{2p+3})$.  Therefore since $\|U-U_p\|\in O(\dt^{2p+1})$ there exist operators $C$ and $E$ that are independent of $\dt$, such that
\begin{equation}\label{eq:suzgeneraln5}
U_p(\initt+\dt,\initt)-U(\initt+\dt,\initt)=C(\initt)\dt^{2p+1}+E(\initt)\dt^{2p+2}+
O(\dt^{2p+3}).
\end{equation}
We then use the above equation to write $U_{p+1}$ as
\begin{align}
\biggr(U(\initt+\dt,\initt+[1-s_p]\dt)&+C(\initt+[1-s_p]\dt)(s_p\dt)^{2p+1}+\cdots \biggr)\times\cdots\nonumber\\
&\times \biggr(U(\initt+s_p\dt,\initt)+C(\initt)(s_p\dt)^{2p+1}+\cdots \biggr).
\end{align}
Since $\{H_j\}$ is $2(p+1)$-smooth, and since $H=\sum_{j=1}^mH_j$,  it follows that $U_{p+1}$ is differentiable $2(p+1)$ times.  Then since $U$ is differentiable $2(p+1)$ times it follows from Taylor's theorem that $C$ is differentiable, and hence we can Taylor expand each $C$ in this formula in powers of $\dt$ to lowest order. By doing so and by defining $\tilde{E}(\initt)$ to be
the sum of all the terms  that are proportional to $\dt^{2p+2}$ in this expansion we find that
\begin{equation}\label{eq:suzgeneraln8}
U_{p+1}(\initt+\dt,\initt)=U(\initt+\dt,\initt)+[4s_p^{2p+1}+[1-4s_p]^{2p+1}]C(\initt)\dt^{2p+1}+\tilde{E}(\initt)\dt^{2p+2}+O(\dt^{2p+3}).
\end{equation}
Then we see that if $s_p=(4-4^{1/(2p+1)})^{-1}$, then the terms of order ${2p+1}$ in the above equation vanish.  Hence the error invoked using $U_{p+1}$ instead of $U$ is $O(\dt^{2p+2})$ with this choice of $s_p$.

Next we show that $\tilde{E}(\initt)= 0$ using reasoning that is
similar to that used by Suzuki in
his proof of his recursive method for the case where $H$ is a constant operator
\cite{suzuki:physa,suzuki:mathphys}. Because $U_{p+1}$ is symmetric it follows from Definition \ref{def:symmetric} that
\begin{align}
\openone&=U_{p+1}(\initt,\initt+\dt)U_{p+1}(\initt+\dt,\initt)&\nonumber\\
&=\left(U(\initt,\initt+\dt)
+\tilde{E}(\initt+\dt)\dt^{2p+2} \right)\left(U(\initt+\dt,\initt)
+\tilde{E}(\initt)\dt^{2p+2} \right)+O(\dt)^{2p+3}\nonumber\\
&=\openone+\left[ U(\initt,\initt+\dt)\tilde{E}(\initt)+\tilde{E}(\initt+\dt)
U(\initt+\dt,\initt)\right]\dt^{2p+2}+O(\dt)^{2p+3}.
\end{align}
This equation is only valid if $\left[ U(\initt,\initt+\dt)
\tilde{E}(\initt)+\tilde{E}(\initt+\dt)U(\initt+\dt,\initt)\right]\in O(\dt)$.

{We then show that $\tilde{E}$ is zero by taking the limit of the
above equation as $\dt$ approaches zero.  But we need to ensure that $E$ is
continuous to evaluate this limit.  The operator $\tilde{E}$ consists of products of derivatives of elements
from the set $\{H_j\}$, and these derivatives are of order at most $2p+1$.  Then since each
$H_j$ is differentiable $2p+2$ times $\tilde{E}$ is differentiable, and hence it is
continuous.}
Then using this fact it follows that
\begin{equation}
\lim_{\dt\rightarrow 0}\left[ U(\initt,\initt+\dt)\tilde{E}(\initt)+\tilde{E}(\initt+\dt)U(\initt+\dt,\initt)\right]=2\tilde{E}(\initt)=0\nn.
\end{equation}

This implies that the norm of the difference between $U$ and $U_{p+1}$ is proportional to $\dt^{2p+3}$, which concludes our proof of Theorem \ref{thm:suzgeneral}.
\end{proofb}

Now that we have proved that Suzuki's recursive method will generate a higher order decomposition formula from a lower order one if $H$ is the sum of the elements from a sufficiently smooth set $\{H_j\}$, we now show that using the $k^{\text{th}}$ order Lie-Trotter-Suzuki product formula invokes an error that is proportional to $\dt^{2k+1}$ if $\{H_j\}$ is $2k$-smooth.

\begin{corollary}\label{cor:lietrotsuz}
Let $H(u)=\sum_{j=1}^m H_j(u)$ where the set $\{H_j\}$ is $2k$-smooth on the interval $[\initt,\initt+\dt]$ and let $U(\initt+\dt,\initt)$ be the ordered operator exponential generated by $H$. Then if $U_k(\initt,\initt+\dt)$ is the $k^{\text{th}}$ order Lie-Trotter-Suzuki formula, then
\begin{equation}
\|U(\initt+\dt,\initt)-U_k(\initt+\dt,\initt)\|\in O(\dt^{2k+1}).
\end{equation}
\end{corollary}
\begin{proofb}

Our proof of the corollary follows from an inductive argument on $k$.  The validity of the base case can be verified by using Lemma \ref{lem:taylr}.  More specifically, since $\{H_j\}$ is $2k$-smooth on $[\initt,\initt+\dt]$ and since $k\geq 1$, then $H$ is at least three times differentiable on that interval. This means that we can use Lemma \ref{lem:taylr} to say that
\begin{align}\label{eq:ulowest}
U(\initt+\dt,\initt)&=\openone+H(\initt)\dt+[H^2(\initt)+H'(\initt)]\dt^2/2+O(\dt^3).
\end{align}
This expansion is also obtained by Taylor expanding $\exp(H(\initt+\dt/2)\dt)$ to third order, so
\begin{equation}\label{ineq1}
\|U(\initt+\dt,\initt)-\exp(H(\initt+\dt/2)\dt)\|\in O(\dt)^3.
\end{equation}
Since $U_1(\initt+\dt,\initt)$ is the Lie-Trotter formula for a constant $H$
equal to $H(\initt+\dt/2)$ it follows that,
\begin{equation}\label{ineq2}
\|U_1(\initt,\initt+\dt)-\exp(H(\initt+\dt/2)\dt)\|\in O(\dt)^3.
\end{equation}
It follows from the above equations and from the triangle inequality that the norm of the difference between $U_1$ and $U$ is at most proportional to $\dt^{3}$.

Since we have shown that $U_1(\initt+\dt,\initt)$ is a symmetric approximation formula whose error is $O(\dt^{3})$, it then follows from Theorem \ref{thm:suzgeneral} and induction, that if $\{H_j\}$ is $2k$-smooth then a symmetric approximation formula whose error is $O(\dt^{2k+1})$ can be constructed from $U_1(\initt+\dt,\initt)$ by applying  Suzuki's recursive method to it $k-1$ times.
\end{proofb}

We have shown in this section that if $\{H_j\}$ is $2k$-smooth on the interval $[\initt,\initt+\dt]$ and if $p\leq k$ then Suzuki's recursive method can be used to create a symmetric decomposition whose error is $O(\dt)^{2p+1}$ out of a symmetric decomposition whose error is $O(\dt)^{2p-1}$.  Then we have used this fact to show that the norm of the difference between $U(\initt+\dt,\initt)$ and the $k^{\text{th}}$ order Lie-Trotter-Suzuki formula is $O(\dt^{2k+1})$.  In the following section we strengthen this result by providing an upper bound on the error invoked by using the $k^{\text{th}}$ order Lie-Trotter-Suzuki product formula.

\section{Error Bounds and Convergence for Decomposition}\label{sec:suzerror}
We showed in Sec.~\ref{sec:suzukianalytic} that if the $k^{\text{th}}$ order Lie-Trotter-Suzuki product formula is used
in the place of the ordered operator exponential of $H$, then an error is incurred that is at most proportional to $\dt^{2k+1}$ if
$H=\sum_{j=1}^m H_j$ and the set of operators $\{H_j\}$ is sufficiently smooth.  We also showed that a sufficient condition for smoothness of the set $\{H_j\}$ is a condition that
we called $2k$-smooth, where this condition is defined is Definition \ref{def:suzsmooth}.
In this section we extend that result by finding upper bounds on the error invoked in using the Lie-Trotter-Suzuki product formula to approximate ordered operator exponentials if $\{H_j\}$ is $\Lambda$-$2k$-smooth.
Unlike the previous section, here we assume that
$\ubound$ is at most one.  This assumption is important because it ensures that our error
bounds are not exponentially large.  Our work can be made applicable to the case where this norm is greater than one by re-normalizing $U$.  We discuss the implications of this in Appendix B.

We first provide in this section an upper bound on the error invoked in using the $k^{\text{th}}$ order Lie-Trotter-Suzuki product formula to approximate the ordered operator exponential $U(\initt+\dt,\initt)$ if $\dt$ is sufficiently short.  We then use this result to upper bound the error if $\dt$ is not short.  More specifically, we show that for every $1\leq\epsilon>0$ and $\dt>0$ there exists an integer $r$ such that
\begin{equation}\label{eq:suzlongapprox}
{\left\|U(\initt+\dt,\initt)-\prod_{q=1}^rU_k\left(\initt+q\dt/r,\initt+(q-1)\dt/r \right)  \right\|}\leq \epsilon,
\end{equation}
if $\{H_j(u)\}$ is $2k$-smooth on the interval $[\initt,\initt+\dt]$.
Finally by
 multiplying the number of exponentials in each $k^{\text{th}}$ order Lie-Trotter-Suzuki product formula by $r$, we find the number of exponentials used in the product in \eqref{eq:suzlongapprox}.  We then use this result to prove Theorem \ref{thm:nexp}.

Our upper bound on the error invoked by using a single $U_k$ to approximate the ordered operator exponential $U(\initt+\dt,\initt)$ is given in Theorem \ref{thm:errbd}.  Before stating Theorem \ref{thm:errbd} we first define the following terms.  Since the $k^{\text{th}}$ order Lie-Trotter-Suzuki product formula is a product of $2m5^{k-1}$ exponentials, we can express this product as $\prod_{c=1}^{2m5^{k-1}}\exp(H_{j_c}(\initt_c)\dt_c)$.  We then
use this expansion to define the following two useful quantities.
\begin{definition}\label{def:Q}
We define $q_{c,2k}\equiv\frac{\dt_c}{\dt}$ and also define $Q_{k}\equiv\max_c |q_{c,2k}|$.
\end{definition}
It can be shown that $Q_1=1/2$ and that if $p>1$ then $Q_p=|1-4s_1|\cdots |1-4s_{p-1}|$.  We show in Appendix A that for any integer $p$ that $Q_p\leq 2p/3^p$, implying that
$Q_p$ decreases exponentially with $p$.  We use this definition of $Q_k$ in the following Theorem, that
gives an upper bound on the difference between the ordered operator exponential $U(\initt+\dt,\initt)$, and the $k^{\text{th}}$ order Lie-Trotter-Suzuki product formula $U_k(\initt+\dt,\initt)$.

\begin{theorem}\label{thm:errbd}
Let $H(u)=\sum_{j=1}^m H_j(u)$, let $\{H_j(u)\}$ be $\Lambda$-$2k$-Suzuki-smooth on the interval
$[\initt,\initt+\dt]$,  and let $\ubound\leq 1$. Then if
$2\sqrt 2(5)^{k-1}Q_k\Lambda\dt\le 1/2$, it follows that
\begin{equation}
\|U(\initt+\dt,\initt)-U_k(\initt+\dt,\initt) \|\le 2\left[3(5)^{k-1}Q_k\Lambda\dt\right]^{2k+1}\nonumber,
\end{equation}
where $U_k$ is given in Definition \ref{def:lietrotsuz}.
\end{theorem}

The proof of Theorem \ref{thm:errbd} requires us to first prove two Lemmas before we can
conclude that the theorem is valid.  We now introduce some notation to state these lemmas concisely.
 Since we have assumed that $\{H_j\}$ is $2k$-smooth, Theorem \ref{thm:suzgeneral} implies that the difference between
$U(\initt+\dt,\initt)$ and $U_k(\initt+\dt,\initt)$ is $O(\dt)^{2k+1}$.  Then using this fact, we know that we only need to compare the terms of $O(\dt^{2k+1})$ to bound the difference between $U$ and $U_k$.  We introduce the following
notation to denote only those terms that do not necessarily cancel.
\begin{definition}\label{def:R}
If the operator $A(\dt)$ can be written as
$A(\dt)=\sum_{p=0}^{2k}A_p\dt^{p}+R(\dt)$ where the norm of $R(\dt)$ is
$O(\dt)^{2k+1}$, then we define $\mathbf{R}_{2k}[A(\dt)]$ to be the norm
of $R(\dt)$.
\end{definition}
This definition simply means that $\mathbf{R}_{2k}$ is the error term for a
Taylor expansion to order $2k$.
Then using this definition, it follows from the triangle inequality that the norm of the difference between $U$ and $U_k$ is at most
\begin{equation}\label{eq:rtriangle}
\mathbf{R}_{2k}\left[U(\initt+\dt,\initt)\right]+\mathbf{R}_{2k}\left[U_k(\initt+\dt,\initt)\right].
\end{equation}
Our proof of Theorem \ref{thm:errbd} then follows from \eqref{eq:rtriangle} and upper bounds that we place on $\mathbf{R}_{2k}[U(\initt+\dt,\initt)]$ and $\mathbf{R}_{2k}[U_k(\initt+\dt,\initt)]$.  Our bound on $\mathbf{R}_{2k}[U(\initt+\dt,\initt)]$ follows directly from Lemma \ref{lem:taylr}, but the bound on $\mathbf{R}_{2k}[U_k(\initt+\dt,\initt)]$ does not.  We will provide the latter upper bound in Lemma \ref{lem:suzerror2}, {but first we provide Definition \ref{def:X} and Lemma \ref{lem:suzerror1}}.

\begin{definition}\label{def:X}
Let $k$ be an integer and let $H=\sum_{j=1}^m H_j$ then $U_k(\initt+\dt,\dt)$ can be written as a product of the form $\prod_{c=1}^{2m5^{k-1}}\exp(H_{j_c}(\initt_c)\dt_c)$.  We then define $X_{p}$ for $p<2k$ to be
\begin{equation}\label{eq:xpdef}
X_p\equiv\sum_{c=1}^{2m5^{k-1}}\left\|H_{j_c}^{(p)}(\initt)\right\|\frac{(\initt_c-\initt)^p}{\dt^p}|q_{c,2k}|,
\end{equation}
and for $p=2k$ we define $X_{2k}$ to be
\begin{equation}\label{eq:x2kdef}
X_{2k}\equiv \sum_{c=1}^{2m5^{k-1}}\max_{\tau\in[\initt,\initt+\dt]}\left\|H_{j_c}^{(2k)}(\tau)\right\|\frac{(\initt_c-\initt)^{2k}}{\dt^{2k}}|q_{c,2k}|.
\end{equation}
Here the quantity $q_{c,2k}$ is given in Definition \ref{def:Q}.
\end{definition}

Then using this definition our lemma can be expressed as follows.
\begin{lemma}\label{lem:suzerror1}
Let $H(u)=\sum_{j=1}^m H_j(u)$ and let the set $\{H_j\}$ be $2k$-smooth on the interval
$[\initt,\initt+\dt]$, then the norm of the difference between $U_k(\initt+\dt,\initt)$ and its Taylor series in powers of $\dt$
truncated at order ${2k}$, is bounded above by
\begin{equation}
\mathbf{R}_{2k}\left[\exp\left(\sum_{p=0}^{2k}\frac{X_p}{p!}
\dt^{p+1}\right)\right],
\end{equation}
\end{lemma}

\begin{proofb}
We begin our proof of Lemma \ref{lem:suzerror1} by writing $U_k$ as a product of $2m5^{k-1}$ exponentials and use Taylor's theorem to write $U_k$ as
\begin{equation}\label{eq:ukexpansion}
\prod_{c=1}^{2m5^{k-1}}\exp\left[  \left(\sum_{p=0}^{2k-1}\frac{H_{j_c}^{(p)}(\initt)(\initt_c-\initt)^p}{p!}
+\int_\initt^{\mu_c}H_{j_c}^{(2k)}(s)\frac{(\initt_c-s)^{2k-1}}{(2k-1)!}\mathrm{d}s\right) \dt_c\right].
\end{equation}
We introduce the terms $v_c=(\initt_c-\initt)/\dt$ and $q_{c,k}=\frac{\dt_c}{\dt}$ and use them to write $U_k(\initt+\dt,\initt)$ as
\begin{equation}\label{eq:ukexpansion2}
\prod_{c=1}^{2m5^{k-1}}\exp\left[ \left(\sum_{p=0}^{2k-1}\frac{H_{j_c}^{(p)}(\initt)v_c^p\dt^p}{p!}
+\int_0^{v_c}H_{j_c}^{(2k)}(\initt+x\dt)\frac{(v_c-x)^{2k-1}}{(2k-1)!}(\dt)^{2k} dx\right) |q_{c,k}|\dt\right].
\end{equation}

We now prove the lemma by placing an upper bound on $\mathbf{R}_{2k}[U_k(\initt+\dt,\initt)]$ by expanding this equation in powers of $\dt$, while retaining only those terms of order
${2k+1}$ and higher. As mentioned previously, the lower order terms are irrelevant since Theorem \ref{thm:suzgeneral} guarantees that they cancel.

By expanding the exponentials in \eqref{eq:ukexpansion2}, taking the norm, using the triangle inequality, upper bounding
each of the norms present in the expansion, and collecting terms again, we find that an upper bound on $\mathbf{R}_{2k}[U_k(\initt+\dt,\initt)]$ is
\begin{equation}\label{eq:ukexpansion3}
 \mathbf{R}_{2k}\left(\exp\left[ \sum_{c=1}^{2m5^{k-1}}\left(\sum_{p=0}^{2k-1}\frac{\left\|H_{j_c}^{(p)}(\initt)\right\|v_c^p\dt^p}{p!}
+\max_{\tau\in[\initt,\initt+\dt]}\left\|H_{j_c}^{(2k)}(\tau)\right\|\frac{(v_c\dt)^{2k}}{(2k)!}\right) |q_{c,k}|\dt\right]\right).
\end{equation}
This equation can be simplified by substituting the constants $X_p$ into it.  These constants are introduced in Definition \ref{def:X}.  After this substitution our upper bound becomes
\begin{equation}
\label{eq:upper1}
\mathbf{R}_{2k}\left[\exp\left(\sum_{p=0}^{2k}\frac{X_p}{p!}
\dt^{p+1}\right)\right].
\end{equation}
\end{proofb}

We use Lemma~\ref{lem:suzerror1} to provide an upper bound on the sum
of the norm of all terms in the Taylor expansion of $U_k(\initt+\dt,\initt)$
which are of order ${2k+1}$ or higher.  This bound is given in the
following lemma.

\begin{lemma}\label{lem:suzerror2}
Let $H(u)=\sum_{j=1}^m H_j(u)$ where the set $\{H_j\}$ is $\Lambda$-$2k$-smooth on the
interval $[\initt,\initt+\dt]$ and let $2\sqrt 2(5)^{k-1}Q_k\Lambda\dt\le \frac{1}{2}$.
Then the norm of the difference between $U_k(\initt+\dt,\initt)$ and its Taylor
series in $\dt$ truncated at order ${2k}$, is upper bounded by
\begin{equation}
\mathbf{R}_{2k}[U_k(\initt+\dt,\initt)]\leq2\left(2\sqrt 2(5)^{k-1}Q_k\Lambda\dt\right)^{2k+1}.
\end{equation}
\end{lemma}

\begin{proofb}

To simplify the following discussion we introduce $\Gamma_{2k}$, defined by
\begin{align}\label{eq:gammadef}
\Gamma_{2k}&\equiv\max_{p=0,\ldots,2k}X_p^{1/(p+1)}.
\end{align}
We first find an upper bound on~$\Gamma_{2k}$.  Now, by Definition~\ref{def:X},
\begin{equation}
X_{p} \leq
\sum_{c=1}^{2m5^{k-1}} |q_{c,k}|  \left(\frac{\mu_c-\initt}{\dt}\right)^p
\max_{\tau\in[\initt,\initt+\dt]}  \left\|H_{j_c}^{(p)}(\tau)\right\|.
\end{equation}
In Appendix~A, we show that $q_{c,k} \leq Q_k \leq
\frac{2k}{3^k}$ for all $c$, where $Q_k$ is an upper bound on the
$q_{c,k}$ given in Definition~\ref{def:Q}.  For a $2k^{\text{th}}$ order
Lie-Trotter-Suzuki product formula, $\frac{\mu_c-\initt}{\dt} \leq 1$.  Plugging
these two bounds into the above inequality and using the fact that each element of $\{H_j\}$ occurs $2(5^{k-1})$ times
in $U_k$ yields,
\begin{equation}
X_{p} \leq 2 (5)^{k-1} Q_k
\sum_{j=1}^{m} \max_{\tau\in[\initt,\initt+\dt]}  \left\|H_{j}^{(p)}(\tau)\right\|.
\end{equation}
Since we assume that $\{H_j\}$ is $\Lambda$-$2k$-smooth, then
$\sum_{j=1}^m\|H_{j}^{(p)}(\tau)\| \leq \Lambda^{p+1}$, so
$X_{p}^{1/(p+1)}
\leq \big(2(5)^{k-1}Q_k\big)^{1/(p+1)} \Lambda$.  We also show in
Appendix~A that $Q_k \geq \frac{1}{2}
\frac{1}{3^{k-1}}$, implying that $2(5)^{k-1}Q_k \geq 1$, and thus
that $X_{p}^{1/(p+1)} \leq 2(5)^{k-1}Q_k \Lambda$.  Since this upper
bound holds for all $0 \leq p \leq 2k$, we conclude that
\begin{equation}\label{eq:gammaupperbound}
\Gamma_{2k} \leq 2(5)^{k-1}Q_k \Lambda.
\end{equation}

We begin the main inequality in Lemma~\ref{lem:suzerror2},
by expanding $\mathbf{R}_{2k}\left[\exp\left( \sum_{p=0}^{2k}\frac{X_p}{p!}
\dt^{p+1}\right)\right]$ in powers of $\dt$ and using
 $\frac{1}{p!} \leq \frac{(\sqrt 2)^{p+1}}{p+1}$ and $X_p
\leq \Gamma_{2k}^{p+1}$. Writing the resulting expansion as an exponential we find that

\begin{equation}
\label{eq:sqrt2bound}
\mathbf{R}_{2k}\left[\exp\left( \sum_{p=0}^{2k}\frac{X_p}{p!}
\dt^{p+1}\right)\right]\leq\mathbf{R}_{2k}\left[\exp\left( \sum_{p=0}^{2k}\frac{(\sqrt 2
\Gamma_{2k}\dt)^{p+1}}{p+1} \right)\right].
\end{equation}
It then follows that the right hand side of
the above expression is upper bounded by
\begin{equation}
\mathbf{R}_{2k}\left[\exp\left(
\sum_{p=0}^{\infty}\frac{(\sqrt{2}\Gamma_{2k}\dt)^{p+1}}{p+1} \right)
\right].
\end{equation}
Using the Taylor expansion of $\ln(1-x)$, we rewrite this as
\begin{equation}
\label{eq:r2klast}
\mathbf{R}_{2k}\left[\exp\left( -\ln(1-\sqrt{2}\Gamma_{2k}\dt)
\right)\right]
= \mathbf{R}_{2k}\left[\frac{1}{1-\sqrt{2}\Gamma_{2k}\dt} \right]
= \sum_{p=2k+1}^\infty (\sqrt{2}\Gamma_{2k}\dt)^p.
\end{equation}
Provided $\sqrt{2}\Gamma_{2k}\dt\leq \frac{1}{2}$, this is upper
bounded by $2(\sqrt 2\Gamma_{2k}\dt)^{2k+1}$.  Plugging inequality
(\ref{eq:gammaupperbound}) into Eq. \eqref{eq:r2klast} then gives that
\begin{equation}
\mathbf{R}_{2k}\left[\exp\left( \sum_{p=0}^{2k}\frac{X_p}{p!}
\dt^{p+1}\right)\right]
\;\leq\; 2\left(2 \sqrt 2 (5)^{k-1}Q_k \Lambda \dt\right)^{2k+1}.
\end{equation}
The lemma follows by applying Lemma~\ref{lem:suzerror1}.
\end{proofb}

Now that we have proven Lemma \ref{lem:suzerror2} we have an upper bound on $\mathbf{R}_{2k}[U_k(\initt+\dt,\initt)]$.  We now use this upper bound to prove Theorem \ref{thm:errbd}.

\begin{proofof}{Theorem \ref{thm:errbd}}

Our proof of Theorem \ref{thm:errbd} begins by recalling the fact that
\begin{equation}
\mathbf{R}_{2k}\left[U(\initt+\dt,\initt)-U_k(\initt+\dt,\initt)\right]\leq \mathbf{R}_{2k}\left[U(\initt+\dt,\initt)\right]+\mathbf{R}_{2k}\left[U_k(\initt+\dt,\initt)\right].\label{eq:rudecomp}
\end{equation}
We then place an upper bound on $\mathbf{R}_{2k}\left[U(\initt+\dt,\initt)\right]$ using Lemma \ref{lem:taylr}.  Using the notation of Lemma \ref{lem:taylr} we write the Taylor series of $U(\initt+\dt,\initt)$ as $\sum_p T_p\dt^{p}/p!$.  We then use the assumption that $\|U(\initt+\dt,\initt)\|$ is less than one, to show from Lemma \ref{lem:taylr} that $\mathbf{R}_{2k}\left[U(\initt+\dt,\initt)\right]$ is at most
\begin{equation}
\frac{\max_{u\in[\initt,\initt+\dt]}\|T_{2k+1}(u)\|\dt^{2k+1}}{(2k+1)!}.
\end{equation}
Using the recursive relations in Lemma \ref{lem:taylr} it follows that $T_{2k+1}$ can be written as
a sum of $(2k+1)!$ terms that are each products of $H$ and its derivatives. Then since $\{H_j:j=1,\dots, m\}$ is $\Lambda$-$2k$-smooth and $H=\sum_{j=1}^m H_j$, it follows from Definition \ref{def:lambdasuzsmooth} that $\|H^{(p)}(u)\|\leq \Lambda^p$ for all $u$ in the interval $[\initt,\initt+\dt]$.  It can then be verified that each term in $T_{2k+1}$ must have a norm that is less than $\Lambda^{2k+1}$.  Therefore it follows that $\|T_{2k+1}\|\leq (2k+1)!\Lambda^{2k+1}$ and hence
\begin{equation}\label{eq:thm3proof}
\mathbf{R}_{2k}\left[U(\initt+\dt,\initt)\right]\leq {(\Lambda\dt)^{2k+1}}.
\end{equation}

Using Eq. \eqref{eq:rudecomp} and Lemma \ref{lem:suzerror2} we see that if $2\sqrt{2}(5)^{k-1}Q_k\Lambda\dt\leq 1/2$ then an upper bound on the sum of  $\mathbf{R}_{2k}\left[U(\initt+\dt,\initt)\right]$ and $\mathbf{R}_{2k}\left[U_k(\initt+\dt,\initt)\right]$ is
\begin{align}
(\Lambda\dt)^{2k+1}+2[2\sqrt 2(5)^{k-1}Q_k\Lambda\dt]^{2k+1}.
\end{align}
We then replace this upper bound with the following simpler upper bound
\begin{equation}
2[3(5)^{k-1}Q_k\Lambda\dt]^{2k+1}.\label{eq:suzerrorfinal}
\end{equation}
This is the claim in Theorem \ref{thm:errbd}, and hence we have proven the theorem.
\end{proofof}

The error bound in Theorem \ref{thm:errbd} is vital to our remaining work,
because it provides us with an upper bound on the error invoked by
approximating an ordered operator exponential by $U_k(\initt+\dt,\initt)$ if $\dt$ is
 short. We will now show a method to devise accurate approximations to the ordered operator exponential $U(\initt+\dt,\initt)$ even if $\dt$ is not short.
 Our approach is similar to that used by Berry et al. in \cite{sanderssim} and that used by Suzuki in \cite{suzuki:commmathphys}; we split the ordered exponential into
 a product of ordinary exponentials, each of which has a short duration.  However to do so we need to present a method to relate the error invoked by using one $U_k$ to the error invoked by using a product of them.  This result is provided in the following lemma.

\begin{lemma}\label{lem:producterror}
If $\|A_{p}-B_{p}\|\leq \delta/P$ where $\delta$ is a positive number less than $1/2$ and $\|A_{p}\|\leq 1$ for every $p\in \{1,2,\cdots,P\}$ then the product $\|\prod_{p=1}^PA_{p}-\prod_{p=1}^PB_{p}\|\leq 2\delta$.
\end{lemma}
\begin{proofb}

Our proof begins by assuming that there exists some integer $q$ such that
\begin{equation}
\left\|\prod_{p=1}^qA_{p}-\prod_{p=1}^qB_{p}\right\|\leq \frac{q\delta \left(1+\frac{\delta}{P} \right)^{q-1}}{P}.
\end{equation}
We then prove Lemma \ref{lem:producterror} by using induction on $q$.  The proof of the base case follows from $\|A_{p}-B_{p}\|\leq \delta/P$.
We then begin to prove the induction step by noting that from $\|A_{p}-B_{p}\|\leq \delta/P$ there exists an operator $C$ with norm
 at most one, such that $B_{q+1}=A_{q+1}+(\delta/P)C$.  Then by making this substitution and using the triangle inequality it follows that
\begin{align}\label{eq:proderror1}
\left\|\prod_{p=1}^{q+1}A_{p}-\prod_{p=1}^{q+1}B_{p}\right\|&\leq\left\|A_{q+1}\left(\prod_{p=1}^{q}A_{p}-\prod_{p=1}^{q}B_{p}\right)\right\|+(\delta/P)\left\| C\prod_{p=1}^qB_{p}\right\|.
\end{align}
Then because $\|A_{p}\|\leq 1$ and $\|B_{p}\|\leq 1+\delta /P$ it can be verified using our induction hypothesis that {the left hand side of} Equation \eqref{eq:proderror1} is bounded above by
\begin{equation}
\frac{(q+1)\delta \left(1+\frac{\delta }{P} \right)^{q}}{P}.
\end{equation}
This proves our induction step, and so it follows that $\|\prod_{p=1}^PA_{p}-\prod_{p=1}^PB_{p}\|\leq \delta(1+\delta /P)^{P-1}$ by using induction on $q$ until $q=P$.  The proof of the Lemma then follows from the fact that if $\delta \leq 1/2$ then $(1+\delta /P)^{P-1}\leq 2$.
\end{proofb}

Using Lemma \ref{lem:producterror} we can now place an upper bound on the
error for decompositions with longer $\dt$.

\begin{lemma}\label{lem:longerror}
If $H(u)=\sum_{j=1}^m H_j(u)$ is $2k$-Suzuki-smooth on the interval
$[\initt,\initt+\dt]$, and $\ubound\leq 1$ and
$\epsilon\le 3Q_k(5)^{k-1}\Lambda\dt$, where $Q_k$ is given in Definition \ref{def:R}, and the positive integer $r$ is greater than
\begin{equation}\label{eq:rval}
\frac{2( 3Q_k(5)^{k-1}\Lambda\dt)^{1+1/2k}}{\epsilon^{1/2k}}
\end{equation}
then we obtain that
\begin{align}
\left\|U(\initt+\dt,\initt)-\prod_{q=1}^{r}U_{k}(\initt+q\dt/r,\initt+(q-1)\dt/r)\right\|\le \epsilon,
\end{align}
where $U_k$ is the $k^{\text{th}}$ order Lie-Trotter-Suzuki product formula, which we introduced in Definition \ref{def:lietrotsuz}.
\end{lemma}

\begin{proofb}

Using the bound $\epsilon\le 3Q_k(5)^{k-1}\Lambda\dt$, we find
using Eq. \eqref{eq:rval} that $3Q_k(5)^{k-1}\Lambda\dt/r\le 1/2$.
Hence we can use Theorem \ref{thm:errbd} to obtain, for each $q=1,\ldots,r$,
\begin{equation}
\left\|U(\initt+q\dt/r,\initt+(q-1)\dt/r)-U_k(\initt+q\dt/r,\initt+(q-1)\dt/r)\right\|
\le 2(3Q_k(5)^{k-1}\Lambda\dt/r)^{2k+1}.
\end{equation}
We then re-write this bound as,
\begin{equation}
2(3Q_k(5)^{k-1}\Lambda\dt/r)^{2k+1}=\frac{2}{r}\frac{(3Q_k(5)^{k-1}\Lambda\dt)^{2k+1}}{r^{2k}}.
\end{equation}
Then from \eqref{eq:rval} we can see that, because $r\geq 2( 3Q_k(5)^{k-1}\Lambda\dt)^{1+1/2k}/{\epsilon^{1/2k}}$ it follows that
\begin{equation}
2(3Q_k(5)^{k-1}\Lambda\dt/r)^{2k+1}\leq\frac{2}{r}\left(\frac{\epsilon}{2^{2k}}\right).
\end{equation}
Then since $k\geq 1$ it follows that
\begin{equation}
\left\|U(\initt+q\dt/r,\initt+(q-1)\dt/r)-U_k(\initt+q\dt/r,\initt+(q-1)\dt/r)\right\|
\le \frac{\epsilon}{2r}.
\end{equation}
Then since both $\epsilon$ and $\ubound$ are less than one, the result of this lemma
follows from Lemma \ref{lem:producterror}.
\end{proofb}

Lemma \ref{lem:longerror} shows that if the maximum value of the norm of $U$ is one, then a product of $k^{\text{th}}$ order Lie-Trotter-Suzuki formulae converges to $U$ as $r$ increases.  Furthermore we can also use this result to prove Theorem \ref{thm:nexp} by using the value of $r$ from this Lemma and multiplying it by the number of exponentials in each $U_k$ to find a bound on the number of exponentials that are needed to approximate $U(\initt+\dt,\initt)$. This proof is presented below.

\begin{proofof}{Theorem \ref{thm:nexp}}
It can be verified from the definition of $U_k(\initt+\dt,\initt)$ in Theorem \ref{thm:suzgeneral} that there are at
most $2m5^{k-1}$
exponentials in each $U_k$ and since at most $r$ different $U_k$ in are needed to approximate $U$ within an error of $\epsilon$ then if $\ubound$ and $\epsilon$ are at most one, it follows from Lemma \ref{lem:longerror} that the number
of exponentials used to decompose $U(\initt+\dt,\initt)$ is at most

\begin{equation}
N\leq 2m5^{k-1}r\leq2m5^{k-1}\left\lceil \frac{2( 3(5)^{k-1}\Lambda Q_k\dt)^{1+1/2k}}{\epsilon^{1/2k}} \right\rceil
\end{equation}
if $\epsilon\leq 3Q_k5^{k-1}\Lambda\dt$.
  We then use the upper bound from Appendix A, $Q_k\leq 2k/3^{k}$ and the fact that $(2k)^{1/2k}< 1.5$ to show that,
\begin{equation}\label{eq:nbdfinal}
N\leq\nbound.
\end{equation}
Equation \eqref{eq:nbdfinal} is only valid if $\epsilon\leq 3Q_k5^{k-1}\Lambda\dt$, this bound
can be simplified by using the lower bound on $Q_k$ in \eqref{eq:boundsonQk}. After substituting this
lower bound we then find that $\epsilon \leq (9/10)(5/3)^{k}\Lambda\dt$ also is sufficient to guarantee that \eqref{eq:nbdfinal} is valid, which proves our theorem.
\end{proofof}

Theorem \ref{thm:nexp} provides an upper bound on the number of exponentials that are needed
to decompose an ordered operator exponential using a product of $k^{\text{th}}$ order Lie-Trotter-Suzuki product formula, while guaranteeing that the approximation error is at most $\epsilon$.  In the following section we present a formula that provides a reasonable value of $k$, for a particular set of values for $\epsilon,\Lambda$ and $\dt$.  Furthermore we show that if $\{H_j\}$ is $\Lambda$-$\infty$-smooth and if that formula for $k$ is used, then the number of exponentials used scales near optimally with $\dt$.

\section{Almost Linear Scaling}
\label{sec:suplin}
Reference~\cite{sanderssim} shows that there exist operator
exponentials that, when decomposed into a sequence of $N$
exponentials, require that $N$ scale at least linearly
with~$\dt$ for large $\dt$.  This implies that any decomposition method that does not
use any special properties of the operator being exponentiated, will
also require that $N$ scale at least linearly with~$\dt$.  {We
now show that if there exists a $\Lambda$ such that the set of operators $\{H_j\}$ is
$\Lambda$-$\infty$-smooth on $[\initt,\initt+\dt]$ for every $\dt>0$}, then we can choose $k$ such that
$N$ scales almost linearly in~$\dt$.  Specifically, we show
that $N/\dt$ is sub-polynomial in~$\dt$, i.e., that $\lim_{\dt
\rightarrow \infty} \frac{N}{\dt^{1+d}} = 0$ for all constants
$d>0$, provided that $\ubound\leq 1$.

It follows from Theorem \ref{thm:nexp} and property $\lceil a \rceil \leq a+1$ for any positive $a$, that if $\Lambda\dt>1$ and $\epsilon\leq 1$ then
\begin{equation}
N\leq
3mk\Lambda\dt \left(\frac{25}{3} \right)^{k}\left(\frac{\Lambda\dt}{\epsilon}\right)^{1/2k}.
\end{equation}
We set
\begin{equation}
\label{eq:pickkpseudolinear}
k_0 = \left\lceil \sqrt{\frac{1}{2} \log_{25/3}\left(\frac{\Lambda\dt}{\epsilon}\right)} \,\right\rceil,
\end{equation}
so that $\left(\frac{25}{3} \right)^{k_0} \geq
\left(\frac{\Lambda\dt}{\epsilon}\right)^{1/2k_0}$.  Then
\begin{equation}
\frac{N}{\dt}
\leq 3mk_0\Lambda  \left(\frac{25}{3} \right)^{2k_0}
    = 3mk_0\Lambda  \exp \left(k_0 2\ln\left(\frac{25}{3} \right) \right),
\end{equation}
which is sub-polynomial in~$\dt$, though not poly-logarithmic
in~$\dt$.  In~conclusion, if $H$ is $\Lambda$-$\infty$-smooth
and if we choose $k=k_0$, then the number of exponentials needed to
decompose a ordered operator exponential of $H$ using the $k^{\text{th}}$
order Lie-Trotter-Suzuki formula scales almost linearly in~$\dt$.

This choice of $k_0$ will cause $N$ to scale nearly linearly with $\dt$ if $H$ is~$\infty$-smooth; however if $\{H_j\}$ is only \mbox{$2P$-smooth}
for some positive integer $P$, then we do not expect this because Theorem~\ref{thm:nexp} cannot be used if $k_0> P$.  Hence
a reasonable choice of  $k_0$ is
\begin{equation}
\label{eq:pickkpseudolinear2}
k_0 = \min\left\{P,\left\lceil \sqrt{\frac{1}{2} \log_{25/3}\left(\frac{\Lambda\dt}{\epsilon}\right)} \,\right\rceil\right\}.
\end{equation}

The choice of $k_0$ in \eqref{eq:pickkpseudolinear2} does not allow for near linear scaling of $N$ with~$\dt$, but it does cause
$N$ to be proportional to~$\dt^{1+1/(2P)}$ in the limit of large $\dt$, and causes
$N$ to have the same scaling with~$\dt$ that a {$\Lambda$-$\infty$-smooth $\{H_j\}$ would have
if $\dt$ is sufficiently short}.

In this section we require that~$\ubound\leq 1$ for this near linear scaling result to hold,
but if this inequality does not hold then $U$ can be normalized to ensure that it does, so it may seem that this
result is more general than we claim.  However we note in Appendix B that
the un-normalized error in the decomposition of the un-normalized $U$ can vary exponentially with $\dt$.
As a result we can only guarantee that the value of $N$ needed to ensure that $\|U-\tilde U\|\leq \epsilon$
can be chosen to scale near linearly with $\dt$ if
$\ubound\leq 1$.

\section{Conclusions}\label{sec:conc}
We have presented in this paper a rigorous derivation of Suzuki's recursive method for generating higher order approximations to ordered operator exponentials from a lower order formula.  We have also shown that if $H(u)=\sum_{j=1}^m H_j(u)$ and $\{H_j\}$ is $2k$-smooth, a condition which we define in Definition \ref{def:suzsmooth}, then Suzuki's recursive method can be used to build approximation formulae for the ordered exponential of $H$ while invoking an error that is at most proportional to $\dt^{2k+1}$.  Furthermore we have shown that the $k^{\text{th}}$ order Lie-Trotter-Suzuki product formula has an error at most proportional to $\dt^{2k+1}$ if $H$ is $2k$-smooth.

We have also shown that if $H$ is $\Lambda$-$2k$-smooth, which is a condition that we give in Definition \ref{def:lambdasuzsmooth}, and if $\ubound\leq 1$ then the number of exponentials needed to approximate an ordered operator exponential of $H$ using a product of $k^{\text{th}}$ order Lie-Trotter-Suzuki formulae while invoking a total error of at most $\epsilon$ is bounded above by
\begin{equation*}
N\leq \nbound.
\end{equation*}
Finally we have shown that if $\{H_j\}$ is $\Lambda$-$\infty$-smooth then for long simulations, the number of exponentials needed to approximate ordered operator exponentials of $H$ with an error of at most $\epsilon$ scales close to linearly with $\dt$. Linear scaling has been shown to be optimal \cite{sanderssim}, so in this sense our scheme is nearly optimal.

An important extension of this work will be to provide upper bounds on the number of exponentials used to decompose an ordered operator exponential when the user chooses step size adaptively, rather than the constant sized steps that we use in our derivation.  Choosing these steps adaptively could lead to substantial improvements in the performance of our decomposition for certain ordered exponentials.
\acknowledgements
Nathan Wiebe would like to thank Ali Rezakhani for many helpful discussions.  We would also like to acknowledge the following agencies that have generously funded this research, the MITACS research network, the Canadian Institute for Advanced Research (CIFAR) of which PH is a Scholar, and BCS
is a CIFAR Associate, the Natural Sciences and Engineering Research Council of Canada, the Informatics Circle of Research Excellence, and the Australian Research Council.
\begin{appendix}\label{appendix:bounds}
\section{Derivation of Bounds on $Q_k$}

When proving bounds on the error introduced by our decomposition of an ordered exponential
 in Section \ref{sec:suzerror}, we use the
quantities $Q_k$ defined by
\begin{equation}
Q_{k} = \frac{1}{2} \max\{s_1,|1-4s_1|\}
        \max\{s_2,|1-4s_2|\} \cdots \max\{s_{k-1},|1-4s_{k-1}|\}
\end{equation}
for $k>1$ and $Q_1 = \frac{1}{2}$, where $s_{k}=
\frac{1}{4-4^{1/(2k+1)}}$ for all $k \geq 1$.  We now show that $Q_k$
decreases exponentially in~$k$,
\begin{equation}\label{eq:boundsonQk}
\frac{3}{2} \frac{1}{3^k} \leq Q_k \leq \frac{2k}{3^k}.
\end{equation}
The lower bound follows directly by noting that
$s_k \geq \frac{1}{3}$ for all $k\geq 1$.

Set $a = 2\ln(4)$, which is approximately $2.7726$.  Using that $-x
\geq \ln(1-x)$ for $0 \leq x < 1$, we then have that for $k \geq 1$,
\begin{equation}
\frac{-1}{2k+1} \ln(4) = \frac{-a}{2(2k+1)}
     \geq \ln \left(1 - \frac{a}{2(2k+1)}\right).
\end{equation}
Taking exponentials on either side yields,
\begin{equation}
4^{-1/(2k+1)} \geq 1 - \frac{a}{2(2k+1)}.
\end{equation}
Multiplying by 4 and subtracting 1 on either side gives,
\begin{equation}
4^{2k/(2k+1)} - 1 \geq 3\left(1 - \frac{2a}{3(2k+1)}\right),
\end{equation}
and taking reciprocals then yields,
\begin{equation}
4 s_k - 1 = \frac{1}{4^{2k/(2k+1)} - 1}
\leq \frac{1}{3} \frac{k + \frac12}{k + \frac12 - \frac{a}{3}}
\leq \frac{1}{3} \frac{k + \frac12}{k + \frac12 - 1}.
\end{equation}
Noting that $4s_k - 1 \geq s_k$ since $s_k \geq \frac{1}{3}$, and
using that $s_1 \leq \frac{2}{3}$, we conclude that
\begin{equation}
Q_{k} \leq \frac{1}{2} \frac{2}{3} \frac{1}{3^{k-2}}
           \frac{k-\frac{1}{2}}{\frac{3}{2}}
\leq \frac{2k}{3^k}
\end{equation}
for $k\geq 2$, and by inspection that the inequality $Q_k \leq
\frac{2k}{3^k}$ also holds for $k=1$.

\section{Norms larger than 1}
In this work we have restricted the norm of $U(\lambda,\mu)$ to not exceed
1. This means that the eigenvalues of $H(\lambda)$ can have no positive
real part. In the case where they do, then the analysis can be performed
in the following way. Simply define the new operators
\begin{align}
  H'(\lambda) &= H(\lambda)-\kappa(\lambda)\openone, \\
H'_j(\lambda) &= H_j(\lambda)-(\kappa(\lambda)/m)\openone,
\end{align}
for some $\kappa(\lambda)$ such that the eigenvalues of $H'(\lambda)$ have
no positive real part. Then the result we have given in Theorem
\ref{thm:nexp} will hold for $H'$ and $\{H_j'\}$ (provided we also define
$\Lambda$ in terms of these operators). The difference between $H$ and
$H'$ simply corresponds to a normalization factor; i.e.\
\begin{equation}
U(\lambda,\mu) = U'(\lambda,\mu)e^{\int_{\mu}^{\lambda}\kappa(x)dx}.
\end{equation}
To approximate $U(\lambda,\mu)$ by a series of exponentials, we can simply
use the series to approximate $U'(\lambda,\mu)$, which gives
\begin{align}
U(\lambda,\mu) &\approx e^{\int_{\mu}^{\lambda}\kappa(x)dx} \prod_{i=1}^N
\exp({H_{j_i}'(\lambda_i)\Delta\lambda_i}) \nn
&= K \prod_{i=1}^N \exp({H_{j_i}(\lambda_i)\Delta\lambda_i}),
\end{align}
where $K$ is a normalization correction
\begin{equation}
K =
e^{\int_{\mu}^{\lambda}\kappa(x)dx-\sum_{i=1}^N\kappa(\lambda_i)\Delta\lambda_i}.
\end{equation}

Thus the same series of exponentials can be used, except for a
normalization factor. There is a difference in the final error that can be
obtained, because
\begin{equation}
\left\|U'(\lambda,\mu)-\prod_{i=1}^N \exp({H_{j_i}'(\lambda_i)\Delta\lambda_i})
\right\| \le \epsilon
\end{equation}
implies that
\begin{equation}
\label{eq:exponential}
\left\|U(\lambda,\mu)-K\prod_{i=1}^N \exp({H_{j_i}(\lambda_i)\Delta\lambda_i})
\right\| \le e^{\int_{\mu}^{\lambda}\kappa(x)dx} \epsilon.
\end{equation}
It might be imagined that the \emph{relative} error can be kept below
$\epsilon$ with similar scaling of $N$. That is, that $e^{\int_{\mu}^{\lambda}\kappa(x)dx}$ can be replaced with $\|U(\lambda,\mu)\|$ in \eqref{eq:exponential}.
Unfortunately, that is not the case. The reason is that, due to the
submultiplicativity of the operator norm, $\|U(\lambda,\mu)\|$ can be much
smaller than $e^{\int_{\mu}^{\lambda}\kappa(x)dx}$.

For example, consider the case where $H$ is initially $\sigma_z$ (the
Pauli operator) over an interval $\Delta\lambda/2$, then is $-\sigma_z$
over another interval $\Delta\lambda/2$. Then $U(\lambda,\mu)=\openone$,
and has norm 1, but
$e^{\int_{\mu}^{\lambda}\kappa(x)dx}=e^{\Delta\lambda}$. A small error in
between the two intervals of length $\Delta\lambda/2$ can then yield a
large relative error in the final result. For example, consider the error
$E=e^{i\delta\sigma_y}$. That yields a final result
\begin{equation}
U(\lambda,\mu+\Delta\lambda/2)E U(\mu+\Delta\lambda/2,\mu) = \left[
{\begin{array}{*{20}c}
   {\cos\delta} & {e^{-\Delta\lambda}\sin\delta}  \\
   {-e^{\Delta\lambda}\sin\delta} & {\cos\delta}  \\
\end{array}} \right].
\end{equation}
The error in this result scales as $e^{\Delta\lambda}$, despite the final
norm being small for $U(\lambda,\mu)$.

With the possibility that the norm of $U(\lambda,\mu)$ exceeds 1, our
approach need not give scaling for $N$ that is close to linear in
$\Delta\lambda$. In the lower bound on $N$ in Theorem \ref{thm:nexp}, the
$(1/\epsilon)^{1/2k}$ will be replaced with
\begin{equation}
(1/\epsilon)^{1/2k}e^{\frac 1{2k}\int_{\mu}^{\lambda}\kappa(x)dx}.
\end{equation}
To prevent this term scaling exponentially in $\Delta\lambda$, one would need to take $k$ proportional to $\Delta\lambda$. However, this would result in $(25/3)^k$ scaling exponentially in $\Delta\lambda$. As a result, it does not appear to be possible to obtain subexponential scaling if there is no bound on the norm of $U(\lambda,\mu)$.
\end{appendix}

\end{document}